\documentstyle[11pt]{article}

\begin{document}

\begin{center}
{\Large \bf Trace anomaly for 4D higher derivative scalar-dilaton theory}
\end{center}

\begin{center}
{\large \sc F. Aceves de la Cruz\footnote{\rm E-mail: fermin@ifug2.ugto.mx} 
and V. I. Tkach\footnote{\rm E-mail: vladimir@ifug3.ugto.mx}}\\
~~\\
{\it Instituto de F{\'\i}sica, Universidad de Guanajuato,}\\
{\it A. P. E-143, 37150, Le\'on, Guanajuato, M\'exico.}
\end{center}

\bigskip

\bigskip

{\noindent \sc Abstract:} Trace anomaly for conformally invariant higher
derivative 4D scalar-dilaton theory is obtained by means of calculating
divergent part of one-loop effective action for such system. Its 
applications are briefly mentioned.


\bigskip

\bigskip

\begin{flushleft}
\end{flushleft}

\newpage

The inflationary brane-world scenario at the early universe based on a 
trace anomaly induced effective action in the frame of Randall-Sundrum 
compactification \cite{Lis} has been suggested in \cite{Haw}. However, 
before developing this scenario is necesary to calculate conformal anomaly 
of brane matter.

Conformally invariant theories are very interesting objects.  On the
classical level, conformal symmetry is manifested in the fact that the
stress-energy tensor for the system is traceless. But at the quantum
level, after applying the regularization and (or) renormalization, such
symmetry is broken. This fact leads to so-called Weyl, conformal or trace
anomaly. Once the symmetry is broken, the next step is to integrate this
anomaly for finding a finite (non-local) effective action and then add it
to classical gravity for describing the effects emerging from quantum
theory. 

So, it is well-known that conformally invariant field theories give
rise to trace anomaly at quantum level \cite{Bir}. For the 4D higher
derivative scalar field theory it was found that trace anomaly can be
obtained from the following action
\[
S = \int{d^4x}\sqrt{-g}\varphi\nabla^4\varphi
\]
where $\nabla^4 \equiv \Box^2 -2R^{\mu\nu}\nabla_{\mu}\nabla_{\nu} +
(2/3)R\Box - (1/3)(\nabla_{\mu}R)\nabla^{\mu}$, $\Box \equiv
\nabla_{\mu}\nabla^{\mu}$ and\footnote{In this work we shall set $c$, 
$\hbar = 1$} $\varphi$ with dimension $l^0$; $\nabla^4$ has the unique
conformally covariant structure for a fourth-order differential operator,
and it is self-adjoint \cite{Rie}. 

It is interesting that conformal anomaly can be found also in the case of
external gravity-dilaton background: Nojiri and Odintsov found a similar
situation for 2D or 4D conformally invariant scalar-dilaton system 
\cite{Od1}. 

In this work we search for the trace anomaly for 4D higher derivative
scalar-dilaton theory by calculating the divergent part of the effective
action for such theory.

Thus, first we must calculate trace anomaly. For this, it is helpful to
remember that effective action is constructed in such a way that its
variational derivative with respect to metric tensor give us the mean
value of stress-energy tensor for the conformally invariant system which
is under consideration. Thus, the form of trace anomaly corresponds with
the divergent part of such effective action \cite{Bir}. Then, we calculate
one-loop effective action for theory (\ref{eq1}) and substract from it its
divergent part. 

The action for higher derivative dilaton coupled scalar is
\begin{equation}
S = \int{d^4x}\sqrt{-g}f(\sigma)\varphi\nabla^4\varphi,
\label{eq1}
\end{equation}
where $\varphi$ is the scalar field, $\sigma$ is the dilaton field and 
there is non-minimal coupling between scalar and dilaton
fields. The self-interaction of dilaton is not considered here. 

We shall calculate the one-loop efective action using the background
fields (i. e. classical fields). For this, we must split all fields in the
action (\ref{eq1}) into their classical and quantum parts, according to
the rule $\sigma \rightarrow \tilde{\sigma}$, $g_{\mu\nu} \rightarrow
\tilde{g}_{\mu\nu}$ and $\varphi \rightarrow \varphi + \tilde{\varphi}$,
where the fields with tilde are taken as being classical. 

One-loop effective action is given by the expression \cite{Od2}
\[
\Gamma^{(1)} = 
\frac{i}{2} 
\mbox{Tr}\ln{\frac{\delta^2S[\varphi]}{\delta\varphi(x)\delta\varphi(y)}}.
\]
Applying this to (\ref{eq1}) we get
\[
\Gamma^{(1)} = \frac{i}{2}\mbox{Tr}\ln{H}
\]
where 
\begin{equation}
H \equiv \Box^2 + L^{\mu}\nabla_{\mu}\Box - 
V^{\mu\nu}\nabla_{\mu}\nabla_{\nu} - N^{\mu}\nabla_{\mu} + U
\label{eq2}
\end{equation}
and we have adopted the notations
\begin{eqnarray}
&&L^{\mu} = 2\frac{(\nabla^{\mu}f)}{f}, \nonumber\\ 
&&V^{\mu\nu} = \left(\frac{(\Box 
f)}{f} - \frac{2}{3}R\right)g^{\mu\nu} + 2\left(R^{\mu\nu} + 
\frac{(\nabla^{\mu}\nabla^{\nu}f)}{f}\right), \nonumber\\
&&N^{\mu} = 2\frac{(\nabla^{\mu}\Box f)}{f} + 
\frac{1}{3}(\nabla^{\mu}R) - \frac{2}{3}\frac{(\nabla^{\mu}f)}{f}R, 
\label{eq3}\\ 
&&U = \frac{(\Box^2f)}{f} + 
\frac{1}{6}\frac{(\nabla^{\mu}f)}{f}(\nabla_{\mu}R) + 
\frac{(\nabla^{\mu}\nabla^{\nu}f)}{f}R_{\mu\nu} - \frac{1}{3}\frac{(\Box 
f)}{f}R. \nonumber
\end{eqnarray} 

The method for calculating the divergences of one-loop effective action is 
based on the universal-trace method \cite{Bar}. The divergent part of 
effective action given by the trace-log of operator (\ref{eq2}) has 
been calculated in \cite{Od2, Od3} and it is
\begin{eqnarray}
\mbox{Tr}\ln{H} &=& \frac{2i}{\varepsilon}\mbox{Tr}\left\{-U + 
\frac{1}{4}L^{\mu}N_{\mu} + \frac{1}{6}L^{\mu}\nabla_{\mu}V - 
\frac{1}{6}L^{\mu}\nabla^{\nu}V_{\mu\nu} -\right. \nonumber\\ 
&&- \left. \frac{1}{24}VL^{\mu}L_{\mu} - 
\frac{1}{12}V^{\mu\nu}L_{\mu}L_{\nu} + \frac{1}{2}P^2 + 
\frac{1}{12}S_{\mu\nu}S^{\mu\nu} + \right.\nonumber\\
&&+ \left.\frac{1}{6}R_{\mu\nu}R^{\mu\nu} - \frac{1}{24}R^2 + 
\frac{1}{12}RV - \frac{1}{6}R^{\mu\nu}V_{\mu\nu} + \right.\nonumber\\
&&+ \left.\frac{1}{48}V^2 + \frac{1}{24}V^{\mu\nu}V_{\mu\nu} + 
\frac{1}{60}F - \frac{1}{180}G\right\} 
\label{eq4} 
\end{eqnarray}
where $F \equiv C_{\mu\nu\rho\tau}^2$ is the square of the conformal Weyl 
tensor, $G$ is the Gauss-Bonnet topological invariant and
\begin{eqnarray*}
&&P \equiv \frac{1}{6}R - \frac{1}{2}\nabla^{\mu}L_{\mu} - 
\frac{1}{4}L_{\mu}L^{\mu}, \quad V \equiv V^{\mu}_{\mu},\\
&&S_{\mu\nu} \equiv \nabla_{[\nu}L_{\mu]} + \frac{1}{2}L_{[\nu}L_{\mu]},\\
&&A_{[\mu}B_{\nu]} \equiv \frac{1}{2}(A_{\mu}B_{\nu} - A_{\nu}B_{\mu}).
\end{eqnarray*}
Substituting relation (\ref{eq3}) into (\ref{eq4}) and carrying out 
elementary transformations we obtain
\begin{eqnarray}
\mbox{Tr}\ln{H} &=& 
\frac{2i}{\varepsilon}\mbox{Tr}\left\{-\frac{(\Box^2f)}{f} - 
\frac{1}{3}\frac{(\nabla_{\alpha}f)(\nabla^{\alpha}R)}{f} - 
R_{\mu\nu}\frac{(\nabla^{\mu}\nabla^{\nu}f)}{f} + \right.\nonumber\\
&&+ \left. \frac{1}{6}R\frac{(\Box f)}{f} + 4\frac{(\nabla_{\alpha}f)(\nabla^{\alpha}\Box f)}{f^2} - 
\frac{11}{3}\left(\frac{\nabla_{\alpha}f}{f}\right)^2\frac{(\Box f)}{f} -\right.\nonumber\\ 
&&- \left.\frac{4}{3}R_{\mu\nu}\frac{(\nabla^{\mu}f)(\nabla^{\nu}f)}{f^2} - 
\frac{4}{3}\frac{(\nabla^{\mu}f)(\nabla^{\nu}f)(\nabla_{\mu}\nabla_{\nu}f)}{f^3} + \right.\nonumber\\
&&+ \left.\frac{19}{12}\left(\frac{\Box f}{f}\right)^2 + 
\frac{1}{6}\left(\frac{\nabla^{\mu}\nabla^{\nu}f}{f}\right)^2\right. + 
\nonumber\\
&&+ \left. \frac{1}{60}F - \frac{1}{180}G \right\}.
\label{eq5}
\end{eqnarray}
In the limit $f(\sigma) = 1$ (\ref{eq5}) is reduced to
\[
\mbox{Tr}\ln{H} = 
\frac{2i}{60\varepsilon}\mbox{Tr}\left(F - \frac{1}{3}G\right) 
\]
which (excluding surface terms) is the same result as the one obtained in 
\cite{Rie}.

The effective action is related with the trace of the energy momentum 
tensor $T_{\mu\nu}$ (see e. g. \cite{Bir, Rie, Od2}). According to these 
results, trace anomaly for theory (\ref{eq1}) has the 
same structure as (\ref{eq5}), i. e.
\begin{eqnarray}
T &\equiv& <T^{\mu}{}_{\mu}> \nonumber\\
&=& \mbox{Tr}\left\{-\frac{(\Box^2f)}{f} - 
\frac{1}{3}\frac{(\nabla_{\alpha}f)(\nabla^{\alpha}R)}{f} - 
R_{\mu\nu}\frac{(\nabla^{\mu}\nabla^{\nu}f)}{f} + \right.\nonumber\\
&&+ \left. \frac{1}{6}R\frac{(\Box f)}{f} + 4\frac{(\nabla_{\alpha}f)(\nabla^{\alpha}\Box f)}{f^2} - 
\frac{11}{3}\left(\frac{\nabla_{\alpha}f}{f}\right)^2\frac{(\Box f)}{f} -\right.\nonumber\\ 
&&- \left.\frac{4}{3}R_{\mu\nu}\frac{(\nabla^{\mu}f)(\nabla^{\nu}f)}{f^2} - 
\frac{4}{3}\frac{(\nabla^{\mu}f)(\nabla^{\nu}f)(\nabla_{\mu}\nabla_{\nu}f)}{f^3} + \right.\nonumber\\
&&+ \left.\frac{19}{12}\left(\frac{\Box f}{f}\right)^2 + 
\frac{1}{6}\left(\frac{\nabla^{\mu}\nabla^{\nu}f}{f}\right)^2\right. + 
\nonumber\\
&&+ \left. \frac{1}{60}F - \frac{1}{180}G \right\} 
\label{eq6}
\end{eqnarray}

One can integrate above conformal anomaly for finding the anomaly induced
effective action. Such action may be considered as quantum correction to
FRW equations. It is also interesting that for some specific value of
dilaton (defined to satisfy $T = 0$) the conformal anomaly is absent. This
indicates that such value of dilaton is kind of fixed point where
conformal symmetry may be restored.

\begin{flushleft}
{\large \bf Acknowledgements}
\end{flushleft}

We are grateful to S. D. Odintsov and O. Obreg\'on for their interest in
this work. This research was supported in part by CONACyT under grant
28454E.

\end{document}